\documentclass[12pt,preprint]{aastex}

\begin{document}
\bibliographystyle{apj} \title {Coping with Type~Ia Supernova
``Evolution'' When Probing the Nature of the Dark Energy}

\author {{David Branch\altaffilmark{1,2}}, {Saul
Perlmutter\altaffilmark{2}}, {E.~Baron\altaffilmark{1,2}}, and {Peter
Nugent\altaffilmark{2}}}

\altaffiltext{1}{Department of Physics and Astronomy, University of
Oklahoma, Norman, Oklahoma 73019, USA}

\altaffiltext{2}{Lawrence Berkeley National Laboratory, Berkeley, CA 94720}

\begin{abstract}

Observations of high--redshift Type~Ia supernovae (SNe~Ia) have
provided strong evidence that the dark energy is real, and making
further accurate observations of high--redshift SNe~Ia is the most
promising way to probe the nature of the dark energy.  We discuss one
of the concerns about such a project --- that of coping with SN~Ia
evolution.  We emphasize that SN~Ia evolution differs in an important
respect from the kind of evolution that has foiled some past projects
in observational cosmology, and we outline empirical strategies that
will take it into account.  The supporting role of physical models of
SNe~Ia also is discussed.  Our conclusion is that systematic errors
due to SN~Ia evolution will be small.

\end{abstract}


\clearpage

\section{INTRODUCTION}

A strong empirical case, based on using Type~Ia supernovae (SNe~Ia) to
determine the cosmic distance--redshift relation, has been made that
the expansion of the universe is accelerating, driven by some kind of
``dark energy'' such as a cosmological constant or quintessence
\citep{riess_scoop98,perletal99,riess00}.  The most promising
way to probe the nature of the dark energy is to make further
observations of SNe~Ia to more accurately determine the
distance--redshift relation \citep[e.g.,][]{WG01}.  A frequently
expressed concern is
that SN~Ia evolution --- a systematic variation in the properties of
SNe~Ia as a function of redshift --- may cause an erroneous
distance--redshift relation to be inferred from the data.  In this
paper (which is intended to be sufficiently free of astronomical
jargon to be intelligible to physicists) we are concerned only with
this single issue of SN~Ia evolution; we do not consider related
potential complications such as evolution of interstellar dust.  The
present discussion is partially motivated by what we see as a fairly
common misconception that controlling SN~Ia evolution will require a
thorough understanding of the origin and physics of SNe~Ia.

Note that although absolute distances are needed to determine the
value of the Hubble constant, probing the nature of the dark energy
requires only accurate relative distances.

SNe~Ia have a well deserved reputation for observational homogeneity.
In particular, at their maximum brightness they have similar
luminosities, i.e., they are very good, but not perfect, ``standard
candles'' \citep{brantam92}.  Most SNe~Ia in the current observational
sample have, at the time of maximum brightness, a blue minus visual
apparent magnitude difference $B-V$ (a ``color index'') near 0.0 (like
that of Vega, a hot star of spectral type A0).  A small fraction of
the SNe~Ia are subluminous and redder in color, some being
intrinsically weak events and others simply being highly extinguished
and reddened by interstellar dust.  When the redder events are
eliminated by means of an objective color cut that excludes those
having $B-V>0.2$, the observational dispersion in the peak blue--band
luminosities is only 25 percent \citep{phil99}.  Considering that this
value is affected by interstellar extinction and observational errors,
the intrinsic luminosity dispersion is about 20 percent (corresponding
to 10 percent in distance).  By traditional astronomical standards
this is an excellent standard candle, and as will be discussed below
this is not the best that can be done, even with existing data.

In \S2 we emphasize that SN~Ia evolution differs in an important
respect from the kind of evolution that has defeated some past
endeavors in observational cosmology.  In \S3 we outline two simple
empirical strategies for coping with SN~Ia evolution.  (More
sophisticated statistical strategies may, of course, prove to be
preferable.)  In \S4 the supporting role of SN~Ia models is discussed.
In \S5 we conclude that systematic errors in the distance--redshift
relation due to SN~Ia evolution will be small.

\section{SN~Ia EVOLUTION IS NOTHING BUT PROGENITOR POPULATION DRIFT}

When making counts of the numbers of galaxies per comoving volume
element as a function of redshift, or when using the brightest
galaxies in clusters as standard candles, coping with evolution is
difficult.  This is because the counts and the luminosities depend on
the time (the age of the universe) and therefore, from our
perspective, on the redshift, $z$.  Galaxies and their numbers per
volume element evolve continuously with time in a way that cannot be
controlled by making observations of nearby galaxies.

SN~Ia ``evolution'' is quite different, and the term can be
misleading.  A fundamantal distinction between a SN~Ia and a galaxy is
that the SN~Ia is an {\sl event} that doesn't know the time --- it
just knows the properties of its immediate progenitor star.  Similar
progenitors at different times should produce similar SNe~Ia.

In nearby galaxies, some 14.5~Gyr after the big bang, a wide range of
potential SN~Ia progenitors is available: local galaxies contain
stellar populations of a variety of ages and metallicities
(heavy--element mass fractions).  As we look out to high redshift and
back in time (e.g., back to when the universe was some 4~Gyr old at
$z=1.7$) we should expect a ``progenitor population drift''.  For
example, if any SNe~Ia in nearby galaxies are produced by very old
stellar populations, $\sim10$~Gyr, they will not have counterparts at
high redshift.  We also can expect a slow drift toward lower mean
metallicity at higher redshift.  However, the important point is that
there will be a strong overlap in the nature of the immediate SN~Ia
progenitors at different redshifts.  Any event that appears at one
redshift but has no counterpart at another can be disregarded, however
interesting it may be physically.\footnote{The recent well observed
SN~2000cx \citep{li00cx01} is an example of a very interesting SN~Ia
that does not yet have a known counterpart.}  It follows that
empirical strategies that are effective at correcting (standardizing)
the luminosities of low--redshift SNe~Ia should also correct the
luminosities of high--redshift SNe~Ia and effectively control the
progenitor population drift.  And, because SNe~Ia don't know what time
it is, that's all the SN~Ia evolution there is.

\section{EMPIRICAL STRATEGIES FOR COPING WITH EVOLUTION}

\subsection {Multi--Parameter Luminosity Corrections}

As mentioned above, a color--cut sample of SNe~Ia has a blue--band
luminosity dispersion of 20 percent.  In addition, the peak luminosity
correlates with the time interval during which the luminosity rises
and falls --- the width of the light curve.  The intrinsic luminosity
scatter about the luminosity--width relation is only 10 percent
\citep{phil99}, so a single--parameter luminosity correction can give
relative distances to 5 percent (here we are disregarding the nuisance
of interstellar extinction).  Some form of a luminosity--width
relation has been used in each of the studies of high--redshift SNe~Ia
that have established the existence of the dark energy.  [Although,
because the distributions of the light--curve widths in the
low--redshift and high--redshift SN~Ia samples are statistically
indistinguishable \citep{perletal99} --- thus providing no evidence
for significant evolution out to $z\sim0.5$ --- practically the same
answer is obtained when a luminosity--width relation is {\sl not}
applied.]

Two--parameter luminosity corrections have not yet been applied to
high--redshift SNe~Ia, but they have been introduced for determining
the value of the Hubble constant \citep{tb99,parodi00}.  When the
light curve width and the $B-V$ color index are used to correct the
luminosities of a color--cut sample \citep{hametal96c} of well
observed SNe~Ia whose relative luminosities are well known, the
luminosities are completely corrected to within the quoted
observational errors.  Actually, the luminosities are corrected to
better than that \citep{tripp98}, indicating that the observational
errors were overestimated.  It follows that the two--parameter
correction yields relative distances to within about 3 percent or
better.  From existing data we can't tell how much better.

The multi--parameter luminosity--correction strategy for coping with
SN~Ia evolution is a straightforward extension of the above: 1) Obtain
a great deal of spectroscopic and broad--band photometric data for a
large sample of SNe~Ia in the Hubble flow\footnote{Not so near that
galaxy peculiar velocities contribute significantly to the observed
redshift and not so far that the distance--redshift relation depends
on $\Omega_m$ and $\Omega_\Lambda$.}, where the relative SN~Ia
luminosities are well known; 2) Establish multi--parameter luminosity
corrections as warranted on statistical grounds, and apply them to
SNe~Ia at all redshifts.  If a two--parameter correction gives
relative distances to within 3 percent, it is reasonable to expect
that this strategy will do even better.

\subsection {Comparing Like with Like}

Here the strategy is to have enough good photometric and spectroscopic
data on so many SNe~Ia, well distributed over redshift, to be able to
scrupulously compare only like with like.  Then the assumption is just
that two events that have the same spectroscopic and photometric
properties have the same luminosity.  This sensible assumption is
supported by SN~Ia models (\S4).  If SNe~Ia turn out to be
continuously distributed in parameter space, then in the process of
quantifying what exactly is meant by ``like'', this strategy may
become essentially equivalent to the multi--parameter
luminosity--correction strategy.  But if SNe~Ia break up into discrete
groups in parameter space this strategy will be distinct, because it
will be possible to assign relative distances to events in each group
without needing to establish the relative luminosities of events in
different groups.

\section {THE SUPPORTING ROLE OF SN Ia MODELS}

\subsection {The Current Level of Understanding}

The exact nature of the immediate progenitors of SNe~Ia is not yet
firmly established.  SNe~Ia most likely are produced by carbon--oxygen
white dwarfs that accrete matter from non--degenerate binary companion
stars until they approach the Chandrasekhar limiting mass of 1.4 solar
masses, ignite carbon, undergo thermonuclear instability, and explode
(the single--degenerate scenario).  It now appears that single
degenerate systems should not fail to produce Chandrasekhar--mass
SNe~Ia \citep{nom00,langetal00,bran01a,tgood01}.  It is still
possible, but perhaps unlikely \citep{sainom98}, that some SNe~Ia are
produced by mergers of binary white dwarfs (the double--degenerate
scenario).

In the single--degenerate scenario a nuclear burning front propagates
outward from the center, burning the inner half of the mass to tightly
bound iron--peak isotopes (primarily $^{56}$Ni because of the equality
of the neutron and proton numbers in $^{12}$C and $^{16}$O) and most
of the outer half to intermediate--mass elements such as silicon,
argon, and calcium.  The fusion energy unbinds the white dwarf and
provides the final kinetic energy.  Adiabatic losses cause the ejected
matter to cool rapidly, but reheating by the trapped decay products of
the radioactive $^{56}$Ni (6.2--day half life) and its daughter
$^{56}$Co (77--day half life) powers the light curve.  Nuclear energy
explodes the star, radioactivity makes it shine.  When SNe~Ia are
treated as a one--parameter family, the dominant parameter that
determines the luminosity is $M_\mathrm{Ni}$, the mass of $^{56}$Ni
that is synthesized in the explosion.

In one dimensional nuclear--hydrodynamical explosion models the
velocity of the burning front is parameterized because the flame
propagation is inherently three--dimensional.  Models in which the
velocity always is subsonic are called deflagrations, and those in
which the velocity goes from subsonic to supersonic are called delayed
detonations.   Three--dimensional models are just
beginning to appear \citep{khokh01,hrn00,hnar00}.

From calculations of spectra and light curves of explosion models
\citep[e.g.,][]{nughydro97,hwt98,l94d01} we have quite a good idea of
what a normal SN~Ia ejects: about a Chandraskhar mass, including
$M_\mathrm{Ni} \simeq 0.6$ solar masses, with a kinetic energy of
$10^{51}$~erg so that the velocity at the boundary between the
iron--peak core and the intermediate--mass elements is near 9000
km~s$^{-1}$.  Models that have acceptable light curves and spectra
include the deflagration model W7 of \citet{nomw7}, and the
delayed--detonation models DD4 of \citet{woosdd91} and M36 of
\citet{hofsn94d}.  Models that differ substantially from these have
spectra and light curves that are inconsistent with the observations
of normal SNe~Ia.

Explosion at the Chandrasekhar mass (in the single degenerate
scenario) or at least not very far from it (double degenerate
scenario) is a plausible reason for the impressive homogeneity of
SNe~Ia.  The first--order cause of the diversity among SNe~Ia is a
range in $M_\mathrm{Ni}$.  This may be due to a range in the
white--dwarf carbon--to--oxygen ratio, which in turn may be caused by
a range in the initial (main--sequence) mass of the white dwarf
progenitors.  We know that the diversity among SNe~Ia actually is
multi--dimensional \citep{hat00}.  Other factors that could
contribute to the diversity include the white--dwarf mass accretion
rate, rotation speed, and magnetic field.

\subsection {The Role of Models}

Models support the basic assumption of the like--with--like strategy.
We can make models that have a variety of luminosities and
spectroscopic and photometric properties --- most of which aren't
consistent with observation because the diversity among model SNe~Ia
exceeds the diversity among real SNe~Ia.  {\sl What we don't know how
to make are models that have the same spectroscopic and photometric
properties but different luminosities.}

At present, models are used to indicate which spectroscopic and
photometric observables are likely to be sensitive to the physical
conditions of the ejected matter, and therefore may prove useful for
making multi--parameter luminosity corrections.  As more good data
accumulate, the process of choosing the observables will become purely
empirical. One reason that spectroscopic observables have seldom been
used so far to correct SN~Ia luminosities is that most of the SNe~Ia
with good spectroscopic data are nearer than the Hubble flow, making
it difficult to establish tight correlations with luminosity.  This
situation is expected to improve rapidly in the near future
\citep{nugent_nearby,aldering_nearby}.

Models provide bounds on how much evolution can be expected.  For
example, model calculations indicate that even decreasing the
metallicity all the way to zero cannot provide a substantial shift in
the luminosity--width relation \citep{DHS01}.
Large evolutionary effects are not plausible.

Models provide reassurance that we are not using tools of which we
have no understanding.  For example, the luminosity--width correlation
is understandable \citep{kmh93} in terms of a range in $M_\mathrm{Ni}$
among SNe~Ia.  The higher the value of $M_\mathrm{Ni}$ the higher the
luminosity, and the higher the value of $M_\mathrm{Ni}$ the higher the
temperature, the higher the opacity, the longer the photon diffusion
time, and the broader the light curve.

The ultimate role of models is to help us learn about SNe~Ia.  Using
SNe~Ia to probe the nature of the dark energy does not require SN~Ia
models.  Inferring the physical properties of SNe~Ia from the
observations obviously does.

\section {CONCLUSION}

SN~Ia evolution is unlike the kind of cosmic evolution that presents
such a challenge for some other projects in observational cosmology.
SN~Ia evolution boils down to a modest amount of progenitor population
drift.  Because we expect a large overlap between the properties of
the immediate progenitors of the SNe~Ia at various redshifts, we
expect the empirical strategies for making luminosity corrections to
nearby SNe~Ia to apply generally. Thus SN~Ia evolution should be
controllable to high accuracy.

Probing the nature of the dark energy with SNe~Ia is primarily an
empirical venture, just as establishing the existence of the dark
energy with SNe~Ia was empirical.  If our physical understanding of
SNe~Ia were to change, the existence of the dark energy would remain
established nonetheless.  Consider the analogy with Cepheid variable
stars.  We have a well developed theoretical understanding of the
Cepheid pulsation mechanism, and detailed models that account for the
Cepheid period--luminosity relation (which itself is analogous to the
SN~Ia luminosity--width relation).  Yet when astronomers use Cepheids
to measure the Hubble constant \citep[e.g.,][]{parodi00,WFfinal01},
they use an empirical period--luminosity relation.

Astronomers do not refrain from scrutinizing the radiation from the
cosmic photosphere --- the microwave background radiation --- on the
grounds that \emph{its} progenitor is not known.  (What was going on
before the Planck time?)  Similarly, present uncertainties about the
origin and physics of SNe~Ia need not deter us from exploiting the
radiation from SN~Ia photospheres to learn about the nature of the
dark energy.  SN~Ia evolution is unlikely to be the limiting factor in
our ability to do so.

\bigskip

This material is based upon work supported by the National Science
Foundation under Grants No. AST--9986965 and AST--9731450.

\clearpage

%\bibliography{refs,baron,rte,sn93j,sn1bc,sn1a,snii,cosmology,gals,crossrefs}

\end{document}